\documentstyle{article}
\begin{document}
\title{Gamma Ray Burst Source Statistics in the Presence of Stochastic Errors}
\author{G.S.Bisnovatyi-Kogan
\thanks{This work was supported in part by
RFFI grant 93-02-17106, Astronomy Program of RMSTP
topic 3-169, NSF grant AST 93-20068 and by COSMION} }
\date{}
{\thanks { Space Research Institute, Profsoyuznaya 84/32, Moscow
117810, Russia.}
\maketitle
\begin{abstract}
Selection effects, connected with stochastic errors in
source flux and threshold value
determination are analyzed. Normal and normal logarithmic distributions
of stochastic deviations are considered. These two kind of distributions
produce different effects on the source statistics. Applications to
Gamma Ray Burst
statistics are discussed.
A physical test for checking a close neutron
star model of GRBs is suggested.
\end{abstract}

\section{Introduction}
Statistical investigation of samples of sources is a powerful method
for analyzing their location, origin and evolution. Most important
results were obtained for distant radio sources (Longair, 1966). In
combination with redshift measurements these data permit also to get
estimates for cosmological parameters (Zeldovich and Novikov, 1975).

The BATSE (Meegan et al,1992; Kouveliotou, 1994; Fishman and Meegan, 1995)
curve [$\log N - \log (C/C_{min})$] gives very important information
for making constraints on GRB models and understanding their nature.
Nevertheless, it suffers from different selection effects,
so it seems premature to use it for a critical choice of GRB models.
This curve differs from the straight line with a slope $3/2$,
and the observed isotropy of GRB distribution on the sky is consistent
with the following models:

1) Nearby neutron stars from the disc population.

2) Galactic halo neutron stars.

3) Cosmological model with bursts coming from sources with redshifts
$z \simeq 1 \div 2$.

\noindent
The last two models also explain deviations from the $3/2$ line.
The third one has extreme demands (neutron star
collision): huge energy release in soft gamma region with no
counterpart in
optics or radio. While some theoretical models based on fireball
expansion (M\'esz\'aros \& Rees,1993) seems
available to reproduce the main properties
of GRB, it is questionable to get such a fireball in neutron star
collision (Isern et.al.,1995). There is also a problem of
extended hard gamma - emission,
accompanying the main burst (Hurley et.al.,1994).
The halo model is posing restrictions (Hakkila et al,1995;
Bulik and Lamb, 1995)
to the properties of the
neutron star kinematics (speed at the origin $\sim 10^3$ km/s),
GRB fluence (narrow strip around $10^{42}$ ergs) and location
(at the outer edge of
a sphere with $R \simeq 350$ kpc).

Here we investigate the influence of stochastic errors on the shape of the
[$\log N - \log (C/C_{min})$] curve. Such type of errors have been first
taken into account by Eddington (1913,1940), who considered small
stochastic errors in determination of the observed magnitude of stars.
Fluence or peak flux of GRBs are determined with much larger errors,
related to the uncertainty
in the burst angular direction, its spectral and temporal
variability, background fluctuations.

\section{Source statistics in presence of stochastic errors }
Let $\chi$ be a number density of events, registered by an observer
with a flux $C$. If $\nu$ is a constant burst
frequency per unit volume, $L$ is a constant peak luminosity, $C_x$ is
a variable threshold flux, $g(C_x)$ is a threshold distribution function,
then, following Schmidt et al (1988),
Band (1992), Petrosian (1993) we may write

\begin{equation}
 \label{ref1}
 \chi(C)\,=\,4\pi\nu \int \delta (C-{L \over 4\pi r^2})
\, \theta(C-C_x) \, g(C_x)r^2 \, dr\, dC_x
 \end{equation}

{\bf a) Logarithmic normal distribution.}
Consider first a classical case where threshold effects are neglected.
Actually, Eddington (1913,1940) considered a case with no threshold,
when all sources could be registered. For that case we have instead of
(\ref{ref1})
\begin{equation}
 \label{ref2}
 \chi(C)\,=\,4\pi\nu \int \delta (C-{L \over 4\pi r^2}) \, r^2 \, dr
 \end{equation}
\noindent
which gives

\begin{equation}
 \label{ref3}
 \chi(C)\,=\,\left({L \over 4\pi}\right)^{3/2} {2\pi\nu \over C^{5/2}}
 \end{equation}
\noindent
Following Eddington
(1913), take errors, according to a normal Gaussian distribution
over $\log C$. The observed flux $C$ could be produced by the source,
whose real flux is $C'$ with the probability

\begin{equation}
 \label{ref4}
 {1 \over \Delta \sqrt{\pi}} \exp\biggl[
 -{(\log {C'}-\log {C})^2 \over \Delta^2} \biggr] d(logC')
 \end{equation}
 \noindent
 The observed number density of events $\tilde\chi(C)$ is connected to
 the real distribution $\chi(C')$ by the relation

\begin{equation}
 \label{ref5}
 \tilde\chi(C) \,={1 \over \Delta \sqrt{\pi}} \int_{-\infty}^{\infty}
 \end{equation}
 $$ \times\chi(C')\,
  \exp\biggl[-{(\log {C}'-\log {C})^2 \over \Delta^2}\biggr]
 d(\log {C'})
 $$
 \noindent
 Using (\ref{ref3}) in (\ref{ref5}) we get after integration

\begin{equation}
 \label{ref6}
 \tilde\chi(C) \,=
  \chi(C)\, e^{{25 \over 16} \Delta^2}
 \end{equation}
 \noindent
So, Gaussian logarithmic statistical errors in absence of threshold
does not change the slope of the number density curve, increasing it by
a constant coefficient (Eddington, 1913)

{\bf b) Normal logarithmic distribution with variable threshold.}
For a variable threshold with $g(C_x) \neq \delta (C_0)$
we need to consider a distribution (Schmidt et.al.,1988)
over $\xi={C \over C_x}$

\begin{equation}
 \label{ref7}
 \chi (\xi)=\, 4 \pi \nu \int \delta (\xi C_x-
 {L \over 4 \pi r^2})
 \end{equation}
$$\times \theta(\xi C_x-C_x)\, g(C_x) r^2 \, dr\, dC_x$$

\noindent
which gives after integration

\begin{equation}
 \label{ref8}
 \chi(\xi)\,=\,\left({L \over 4\pi}\right)^{3/2}
 {2\pi\nu A \over \xi^{5/2}} \theta(\xi-1)
 \end{equation}
 with

 $$A=\int_0^{\infty}{g(C_x)\, dC_x \over C_x^{5/2}}$$
\noindent
Consider now a Gaussian logarithmic distribution of errors like (\ref{ref4})
with $\xi$ and $\xi'$ instead of $C$ and $C'$. While it is impossible
to register events under the threshold, we shall use the interval
$1\,<\, \xi'\,< \, \infty$ for possible real values. Then we have
instead of (\ref{ref5})

\begin{equation}
 \label{ref9}
 \tilde\chi(\xi) \,={1 \over B} \int_0^{\infty}
  \chi(\xi')\, \exp{\biggl[-{(\log {\xi'}-\log {\xi})^2 \over \Delta^2}
  \biggr]} d(\log {\xi'})
 \end{equation}
with

$$ B= \int_0^{\infty} \exp{\biggl[
-{(x-\log {\xi})^2 \over \Delta^2}\biggr]} dx$$
 \noindent
Using (\ref{ref8}) in (\ref{ref9}) we get

\begin{equation}
 \label{ref10}
 \tilde\chi(\xi)\,=\,\left({L \over 4\pi}\right)^{3/2}
 {2\pi\nu A\,I \over B \xi^{5/2}}
 \end{equation}
\noindent
Here

\begin{equation}
 \label{ref11}
 B=\Delta(\int_0^{\infty}e^{-z^2}\,dz\,+\int_0^{(\log \xi) /\Delta}
 e^{-z^2}\,dz)
 \end{equation}
 $$=\Delta {\sqrt \pi \over 2} \bigl[1+Erf\left(\log\xi \over
 \Delta \right)\bigr],$$
 $$I=\Delta e^{{25 \over 16} \Delta^2}
 (\int_{{5 \over 4}\Delta}^{\infty}e^{-z^2}\,dz\,+
 \int_{-{5 \over 4}\Delta} ^{{\log \xi \over \Delta}-{5 \over 4}\Delta }
 e^{-z^2}\,dz)$$
 $$=\Delta {\sqrt \pi \over 2} e^{{25 \over 16} \Delta^2}
 \bigl[1+Erf\left({\log \xi \over \Delta}-{5 \over 4}\Delta \right)\bigr].$$
\noindent
where   $Erf(x)={2 \over \sqrt\pi} \int_0^x e^{-z^2}\,dz $.
An observed distribution
$\tilde\Xi(\xi)$
obtained by integration of (\ref{ref10}) over $d\xi$ is written as follows

\begin{equation}
 \label{ref12}
\tilde\Xi(\xi)=\int_{\xi}^{\infty} \tilde\chi(\xi')\,d\xi'
 \end{equation}
 $$=\,\left({L \over 4\pi}\right)^{3/2} 2\pi\nu A\,e^{{25 \over 16} \Delta^2}
 \,\int_{\xi}^{\infty} {D(x) \over x^{5/2}}dx
 $$
\noindent
With
\begin{equation}
 \label{ref13}
 D(x)=
 {1+Erf\left({\log \xi \over \Delta}-{5 \over 4}\Delta \right) \over
1+Erf\left({\log \xi \over \Delta}\right)}
 \end{equation}
\noindent
The function $\log\tilde\Xi$ as a function of $\log\xi$ is represented in
fig.1 for $\Delta=1$. The corresponding curve for $\Delta=0.1$ is
cannot be distinguished from the straight line with slope $3/2$
indicated there. Contrary to the case with no threshold (Eddington,1913),
where stochastic errors do not change the slope of the curve
$[\log N\,-\,\log C]$, stochastic errors in presence of threshold
may considerably decrease the slope of the curve in the vicinity of the
threshold. It happens, because the threshold is acting like a border,
which cannot be crossed by the bursts from both sides. So faint bursts
appear more like stronger ones, and the spreading of stronger bursts is
almost equal in both directions.
\begin{figure}
\vspace{12cm}
\caption{The curve $[\log N - \log C(max)/C(thr)]$ in presence of
stochastic errors, distributed according to normal logarithmic
distribution; 1 - straight line with a slope 3/2, corresponding
to $\Delta=0$; 2 - curve with $\Delta=1$; $C(max)$ is the peak intensity
of the burst; $C(thr)$ is a corresponding threshold value.}\label{Fig.1}
\end{figure}
\hfill\break\indent
{\bf c) Normal distribution.}
While we do not know exactly what law is determining stochastic errors,
let us consider also a normal distribution of errors around the value
$\xi$ itself. While our eye has a logarithmic response to the signal,
some $X$ - ray counters are proportional. May be the true distribution
is even more complicated and does not follow a Gaussian law neither in
logarithms nor in the values themselves.

Assume that the burst, registered with the ratio $\xi=C/C_x$ may in reality
correspond to the ratio $\xi'$ with the probability

\begin{equation}
 \label{ref14}
 \sim\, \exp\biggl[-{(\xi'-\xi)^2 \over \Delta_1^2}\biggr]\, d\xi'
 \end{equation}
\noindent
Then taking into account only the events over the threshold we get
instead of (\ref{ref5})

\begin{equation}
 \label{ref15}
 \tilde\chi(\xi)={1 \over B_1}\int_1^{\infty} \chi(\xi')
 \exp\biggl[-{(\xi'-\xi)^2 \over \Delta_1^2}\biggr]\, d\xi'
 \end{equation}
\noindent
with

\begin{equation}
 \label{ref16}
  B_1=\int_1^{\infty} \exp\biggl[-{(\xi'-\xi)^2
  \over \Delta_1^2}\biggr]\, d\xi'
 \end{equation}
\noindent
Taking into account the (\ref{ref3}) we get

\begin{equation}
 \label{ref17}
 \tilde\chi(\xi)\,=\,\left({L \over 4\pi}\right)^{3/2}
 {2\pi\nu A\,I_1 \over \Delta_1^{3/2} B_1 },
 \end{equation}
\noindent
where the integrals may be expressed as

\begin{equation}
 \label{ref18}
 B_1=\Delta_1 {\sqrt \pi \over 2} \bigl[1+Erf\left(\xi-1 \over
 \Delta_1 \right)\bigr],
 \end{equation}
\begin{equation}
 \label{ref19}
  I_1=\int_{1 \over \Delta_1} ^{\infty} y^{-5/2}
  \exp{\bigl[-\left(y-{\xi \over \Delta_1}\right)^2 \bigr]}\, dy.
 \end{equation}
\noindent
For the distribution we are looking for we get

\begin{equation}
 \label{ref20}
\tilde\Xi(\xi)=\,\left({L \over 4\pi}\right)^{3/2}
{4\pi\nu A \over \sqrt\pi \Delta_1^{5/2}}\,
\int_{\xi}^{\infty} D_1(x)dx
 \end{equation}
\noindent
with

\begin{equation}
 \label{ref21}
 D_1(x)=
 {I_1(x) \over
1+Erf\left({ x-1 \over \Delta_1}\right)}
 \end{equation}
\noindent
For sufficiently large values of $\xi$, when the function $Erf(x)$ is
very close to unity, the expression for $\tilde\Xi(\xi)$ in (\ref{ref20})
may be written as

\begin{equation}
 \label{ref22}
\tilde\Xi(\xi)=\,\left({L \over 4\pi}\right)^{3/2}
{4\pi\nu A \over 3\sqrt\pi \Delta_1^{3/2}}\,
 \biggl[\int_0^{{\xi - 1 \over \Delta_1}}\biggl({\xi \over \Delta_1}
-z \biggr)^{-3/2}
\end{equation}
$$ \times  e^{-z^2} dz
+\int_0^{\infty} \left({\xi \over \Delta_1}+z \right)^{-3/2}\,
e^{-z^2} dz \biggr]$$
\noindent
The plot of $\log\Xi (\xi)$ as a function of $\log\xi$ is represented in
fig.2 for $\Delta_1\,=\,1,\,\,10$, together with the straight line
with the slope 3/2.
\begin{figure}
\vspace{12cm}
\caption{Same as in Fig.1 for normal distribution;
1 - straight line with slope 3/2, corresponding
to $\Delta_1=0$; 2 - curve with $\Delta_1=1$;
3 - curve with $\Delta_1=10$.} \label{Fig.2}
\end{figure}

\section{Discussion}
BATSE consists (Fishman, 1992) of eight detectors, arranged on corners of
the Compton Gamma Ray Observatory (CGRO). Burst registration is done by
large area detectors (LAD), optimized sensitivity and directional response.
The eight panels of LAD are parallel to the eight faces of a regular
octahedron. Since a regular octahedron is comprised of four sets of
parallel intersecting planes, every detected burst will be viewed by
four detectors. LAD are sensitive in the energy range 20-600 keV.
The burst is registered, when $5.5 \sigma $ excess over 17 s background
rate is registered at least by two detectors (Fishman et al, 1992).
The background in LAD in the burst trigger energy range 60-300 keV
varies between approximately 1500 counts/s and 3000 counts/s per
detector during most portion of the orbit and above geographic
latitudes of about $22^{\circ}$ the background increases considerably.

Each event with the ratio $\xi$ of the peak luminosity to a local
background is detected with an error due to the following circumstances.

1) Spectral dispersion of GRBs cannot guarantee that a peak value in the
BATSE spectral region (BSR) is equal to a real one. It may be several
times larger if the region of maximum radiation lay outside BSR.

2) Angular dependence of the detector sensitivity, especially in the
region of a steep dependence around $50^{\circ}$ (Fishman,1992)
imply errors in determination of $\xi$ because of poor angular
localization of the source.

3) Different time duration of bursts lead to nonuniformity of the source
sample, where a same peak luminosity may be related to bursts with a
total flux or average luminosity varying by orders of magnitude. Conversely
very different peak luminosities may correspond to bursts with the same
total flux. This would imitate stochastic errors of the same order.

4) Nonuniformity of GRB detection conditions, when an event may be
registered by 2 or 3 or 4 detectors determines an additional source of
dispersion.

\noindent
So, uncertainty in the $\xi$ value represented by the dispersion equal to
10 threshold levels, described above, does not seems to be overestimated.

For $\Delta_1=10$, corresponding to $\Delta=1$
in the case of logarithmic normal distribution of errors, the shape of the
curve is changed considerably.
The changes become noticeable at $\xi \sim 30$ (see fig.2),
which is much larger, than the average dispersion value $\Delta_1=10$.
In the case of logarithmic dispersion the influence of stochastic errors
starts at $\xi$ approximately equal to the value of the average dispersion
(see fig.1).
Results, represented in fig.1,2 illustrate the large importance of different
stochastic errors in the form of the [$\log N - \log (C/C_{min})$] curve
and are not intended to explain directly the corresponding BATSE curve.

We have not taken into account other kinds of threshold influence,
leading to additional deviations from the 3/2 slope
(Hartman and The, 1993; Lingenfelter, 1995).
Combination of these effects must be taken into account in analyzing
the [$\log N - \log (C/C_{min})$] curve for the BATSE data sample.

Let us note, that stochastic errors from the photon noise, to which
Eddington (1913) applied his calculations, are much smaller than
possible errors in BATSE data caused by above mentioned reasons.
Even in statistical analysis of G- stars, situated uniformly inside the
Galactic disc, the value of $<V/V_{max}>$ which must be equal to 0.5 for
uniform unbiased sample, falls down considerably around two
threshold flux, which is connected most probably with a loss of faint
sources (Harrison et al, 1995).

In view of this
situation it seems preliminary to wipe away a close Galactic origin
for GRBs and all variety of models (see e.g.
Ho et al, 1992) should remain under discussion.

\section{Testing the physical origin of GRBs}
If GRBs are connected
with starquakes on nearby neutron stars (Bisnovatyi-Kogan et al, 1975)
it is worth
(Bisnovatyi-Kogan, 1993)
to monitor close young
pulsars (Geminga) for catching the moment
of the quake and comparing it with GRB search data.
If the hard tail of GRBs (Hurley et al, 1994) is
connected with the excitation of submsec proper oscillations of the neutron
star after starquake, and hard gamma ray emission
in GRBs is produced by the
same mechanism as in radiopulsars (Bisnovatyi-Kogan, 1995), one implicit
test for GRB origin may be suggested.
If starquakes in
radiopulsars lead to the excitation of such oscillations,
they  can lead to the appearance of resonant modes in a frequency
spectrum of radioemission.

The electrical field, generated on the surface of an oscillating neutron star,
is of the order of (Muslimov and Tsygan, 1986)

\begin{equation}
\label{ref23}
E_{osc} \simeq {v \over c} B \approx {\delta R \over R}{v_{ff} \over c} B
\approx {\delta R \Omega_{osc} \over c} B
\end{equation}
\noindent
When the amplitude of oscillations is large enough

\begin{equation}
\label{ref24}
{\delta R \over R} \geq {\Omega_{rot} \over \Omega_{osc}} \simeq
{10^{-4} \over P_{rot}},
\end{equation}
\noindent
this oscillating field could modulate a pair cascade birth, leading to the
appearance of a coherent  high frequency mode in the frequency
spectrum of radioemission.

In old nearby neutron stars (silent ones) we may expect slower rotation
and lower magnetic field than in radiopulsars. These neutron stars
could become pulsars (in hard gamma as well as in radio) only temporally,
for about a few hours after the quake, and produce a GRB, if the electrical
field, induced by oscillations is higher than the threshold field for
pair cascade generation.

The best object for such testing is
the Vela radiopulsar, where strong glitches are observed almost every
year (MacCulloch et.al.,1987).
It is necessary to be able to make a frequency analysis of the
radio data very soon after the quake for checking the existence of resonance
frequencies with periods less then one millisecond.
If the model of close Galactic
GRB with its logical consequence listed above is true, we may expect to
see high frequency resonance oscillations only during a limited period of time
of the order of 90 minutes after the visible glitch. Radio observations
of Vela soon after glitch with high time resolution and accurate frequency
analysis could be more informative than gamma ray
observations (Hartmann et al, 1992).

{\bf Acknowledgements}
The author is very grateful to Prof. M. Schmidt for useful discussions
and for sending him copies of both papers of A. Eddington.

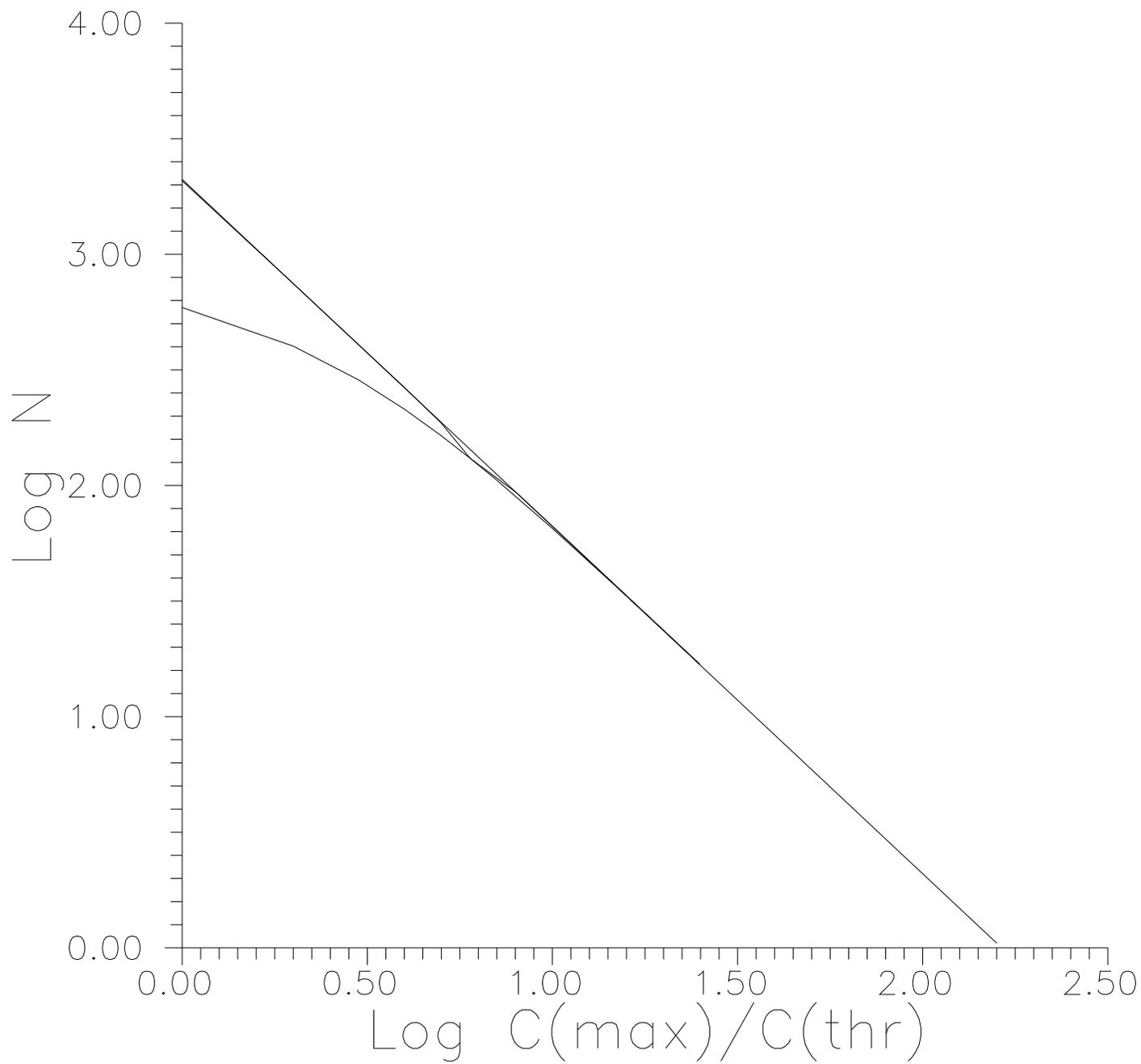
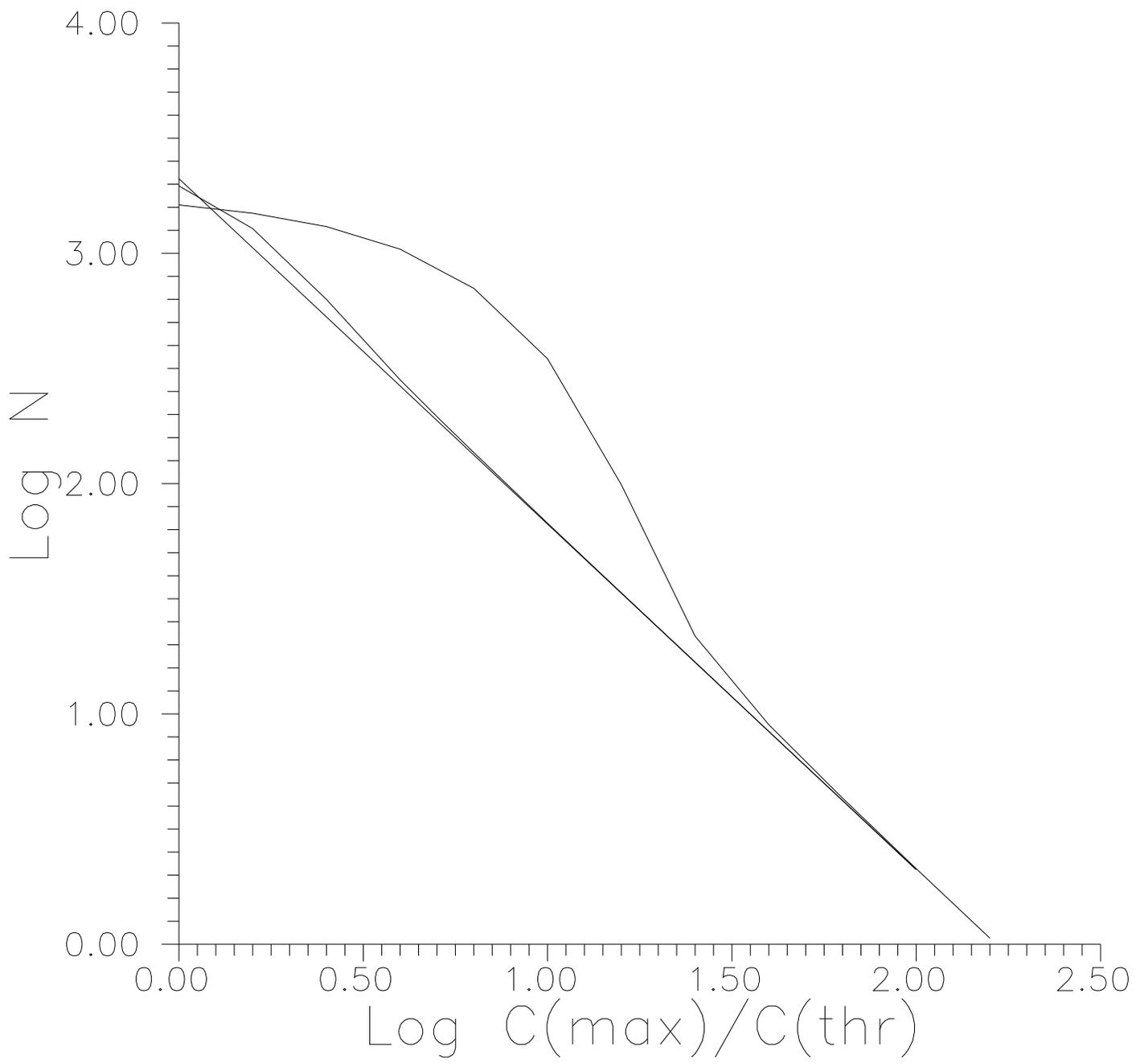
\end{document}